\documentclass[aip,jap,amsmath,amssymb,preprint,
]{revtex4-1}
\usepackage[colorlinks=false,linkcolor=blue]{hyperref}
\usepackage{graphicx}
\usepackage{dcolumn}
\usepackage{bm}
\usepackage{eepic}
\usepackage{textcomp}
\usepackage{subfigure}
\usepackage{xcolor}

\newcommand{\bes}{\begin{subequations}}
\newcommand{\ees}{\end{subequations}}
\newcommand{\bea}{\begin{eqnarray}}
\newcommand{\eea}{\end{eqnarray}}
\newcommand{\ben}{\begin{equation}}
\newcommand{\een}{\end{equation}}
\newcommand{\al}{\alpha}
\newcommand{\ba}{\beta}
\newcommand{\del}{\delta}
\newcommand{\Del}{\Delta}

\begin{document}
\title{A critique on the soliton solutions of $\cal{PT}$-invariant reverse space nonlocal nonlinear Schr\"{o}dinger equation}
\author{S.~Stalin }%
\affiliation{Centre for Nonlinear Dynamics, School of Physics, Bharathidasan University, Tiruchirappalli - 620 024, Tamilnadu, India}%

\author{M.~Senthilvelan}%
\email{velan@cnld.bdu.ac.in}
\affiliation{Centre for Nonlinear Dynamics, School of Physics, Bharathidasan University, Tiruchirappalli - 620 024, Tamilnadu, India}%

\author{M.~Lakshmanan}%
\affiliation{Centre for Nonlinear Dynamics, School of Physics, Bharathidasan University, Tiruchirappalli - 620 024, Tamilnadu, India}%

\begin{abstract}
We point out certain basic misconceptions and incorrect statements given by G\"{u}rses and Pekcan in the recent paper {\bf J. Math. Phys. 59, 051501  (2018)}.  We re-emphasize the soliton solution derived by us earlier in {\bf Phys. Lett. A. 381, 2380 (2017)}  for the reverse space nonlocal nonlinear Schr\"{o}dinger equation is correct and more general and contains the solutions given by G\"{u}rses and Pekcan as special cases. 
\end{abstract}
\maketitle
In this note, we wish to point out certain basic misconceptions and incorrect statements made by G\"{u}rses and Pekcan  in their recent paper \cite{1} on the soliton solutions of space reflection symmetric ($S$-symmetric) nonlocal nonlinear Schr\"{o}dinger (NNLS) equation . Further, we would like to re-emphasize that the soliton solutions for the reverse space  NNLS equation obtained by us in Ref. 2 is correct and more general (both $\cal{PT}$-symmetry preserving/broken cases). The solutions obtained by G\"{u}rses and Pekcan turn out to be special cases of the solutions obtained by us. 

 In Ref. 2, we have constructed one- and two-soliton solutions for the following $\cal{PT}$-symmetric reverse space NNLS equation introduced in Ref. 3, 
\begin{equation}
iq_{t}(x,t)-q_{xx}(x,t)-2q(x,t)q^{*}(-x,t) q(x,t)=0.
\label{1}
\end{equation}
To obtain general soliton solutions of the above equation through a nonstandard bilinearization procedure, we  augmented the evolution equation for the nonlocal field $q^{*}(-x,t)$ which results from the AKNS scheme \cite{4} as
\bea
iq_{t}^{*}(-x,t)+q_{xx}^{*}(-x,t)+2q^{*}(-x,t)q(x,t) q^{*}(-x,t)=0.
\label{2}
\eea 

 In Eq. (\ref{1}), the nonlocal nonlinearity emphasizes the fact that one of the dependent variables is evaluated at $-x$ while the other variable is evaluated at $+x$ simultaneously. This implies that the functions $q(x,t)$  and $q^*(-x,t)$ need not be dependent and they are two independent fields in Eq. (\ref{1}). Due to the above reasons, we treat the nonlinear Schr\"{o}dinger field $q(x,t)$ and the nonlocal field $q^*(-x,t)$ as two independent fields satisfying Eqs. (\ref{1}) and (\ref{2}). Since one is considering the Cauchy initial value problem associated with (1) and (2), it implies that  one can specify $q(x,0)$ and $q^*(-x,0)$ independently and consequently they evolve as the coupled system specified by Eqs. (1) and (2).

 Ablowitz and Musslimani have shown that in Ref. 5 the above reverse space NNLS Eqs. (\ref{1}) and (\ref{2}) posses  the $\cal{PT}$-symmetry  property which has been discussed  widely in the recent literature \cite{6,7}. The $\cal{PT}$-symmetry property ensures that Eq. (\ref{1}) (and Eq. (\ref{2}) as well) is $\cal{PT}$ invariant under the combined transformation of parity ($\cal{P}$): $x\rightarrow -x$ and time reversal ($\cal{T}$): $t\rightarrow -t$ along with $i\rightarrow -i$. Note that an evolution equation admitting certain symmetry  property does not imply that the resultant solution should also exhibit the same symmetry: it may exhibit spontaneously broken symmetry property as well.   

In this situation, if Eq. (\ref{1}) admits a solution which obeys the $\cal{PT}$-symmetry property, that is ${\cal{PT}}\big[q(x,t)\big]=q^*(-x,t)\equiv [q(x,t)]^*_{|_{x\rightarrow-x}}$, then such a solution is called a $\cal{PT}$-symmetry preserving solution. For example, the explicit one soliton solution obtained by Ablowitz and Musslimani exhibits $\cal{PT}$-symmetry \cite{3,4}. Consequently, in this case the function $q^*(-x,t)$ is nothing but the one obtained from the function $q(x,t)$ after taking complex conjugation and a space inversion. We call this case as $\cal{PT}$-symmetry preserving solution. 

On the other hand, if the solution $q(x,t)$ does not obey the above $\cal{PT}$-symmetry property of Eq. (\ref{1}), ${\cal{PT}}\big[q(x,t)\big]\neq q^*(-x,t)$, then we call such a solution as $\cal{PT}$-symmetry broken solution. In this case, the function  $q^*(-x,t)$ need not be parity transformed complex conjugate of  $q(x,t)$. It turns out that Eqs. (\ref{1}) and (\ref{2}) admit both the above types of solutions. 

The above fact ensures that the solutions need not preserve the symmetry while the original evolution equation (reverse space NNLS Eq. (\ref{1}) and (\ref{2})) does. This is akin to spontaneously symmetry breaking solutions, for example  $\cal{P}$-symmetry in $\ddot{x}-\omega_0^2x+\lambda x^3=0$ or $\cal{PT}$-symmetry in $\ddot{x}+kx\dot{x}-\omega_0^2x+\frac{k^2}9{} x^3=0$ or their field versions, see Ref. 8. Note that these equations also admit both symmetry preserving and breaking solutions. In view of the above said reasons, to explore symmetry preserving and non-preserving solutions, it is very much essential to consider the fields $q(x,t)$ and $q^*(-x,t)$ as two independent fields. Once their explicit forms are obtained one can categorize them by imposing or excluding the relation
\ben
q^*(-x,t)=[q(x,t)]^*_{|_{x\rightarrow-x}}.
\label{3}
\een
Note that Eq. (\ref{3}) is an extra requirement not demanded by the Cauchy initial value problem of Eqs. (1) and (2), and so it is not required in general. The situation is similar to a simple time delay equation $\frac{dx}{dt}=-bx+af(x(t-\tau))$, where $a$, $b$ and $\tau$ are constants and $f$ is a nonlinear function. Then the solution $x(t-\tau)$ is not merely $x(t)$ evaluated at $t=t-\tau$ but is much more complicated and chaotic \cite {9} and the initial conditions have to be specified on a line $-\tau\le t\le 0$ and each value of $x(t)$ in this interval evolves independently. 

Considering all the above facts, we have constructed general soliton solutions of reverse space NNLS Eq. (\ref{1}) by solving the later equation along with Eq. (\ref{2}) simultaneously through a nonstandard bilinearization procedure \cite{2}. We have constructed the one-soliton solution of Eq. (\ref{1}) (and (\ref{2})) in the form,   
\bea
&&q(x,t)=\frac{\al_1e^{\bar{\xi}_1}+e^{\xi_1+2\bar{\xi_1}+\del_{11}}}{1+e^{\xi_1+\bar{\xi_1}+\del_1}+e^{2(\xi_1+\bar{\xi_1})+R}}\equiv \frac{\al_1e^{\bar{\xi}_1}}{1+e^{\xi_1+\bar{\xi_1}+\Del}},~e^{\Del}=\frac{-\al_1\ba_1}{(k_1+\bar{k}_1)^2}, \nonumber\\
&&q^{*}(-x,t)=\frac{\ba_1e^{\xi_1}+e^{2\xi_1+\bar{\xi_1}+\Delta_{11}}}{1+e^{\xi_1+\bar{\xi_1}+\del_1}+e^{2(\xi_1+\bar{\xi_1})+R}}\equiv\frac{\ba_1e^{\xi_1}}{1+e^{\xi_1+\bar{\xi_1}+\Del}},
\label{4}
\eea 
and then the two soliton solution. Here, $\xi_1=i k_{1}x-ik_{1}^{2}t+\xi_1^{(0)}$ and $\bar{\xi_1}=i \bar{k_{1}}x+i\bar{k_{1}^{2}}t+\bar{\xi}_1^{(0)}$. In the above solution, all the parameters, namely $\alpha_1$, $\ba_1$, $k_1$, $\bar{k}_1$, $\xi_1^{(0)}$ and $\bar{\xi}_1^{(0)}$ are arbitrary complex constants and in general there exists no relation between them.  From the above solution, one can immediately observe that the functions $q(x,t)$ and  $q^*(-x,t)$ are independent and they satisfy both the Eqs. (\ref{1}) and (\ref{2}) without any restriction among the parameters. The soliton solution given above in (\ref{4}) is in general a $\cal{PT}$-symmetry broken solution, except for special choices of parameters as indicated below. In the general case, the soliton parameters present in the solution (\ref{4}) are not related to each other and they in general do not obey the constraint equation (109) given in Ref. 1.  We have also deduced the $\cal{PT}$-symmetry preserving solution, that is
\bes
\begin{eqnarray}
q(x,t)=-\frac{2(\eta_{1}+\bar{\eta_1})e^{i\bar{\theta}_1}e^{-4i\bar{\eta_{1}}^{2}t}e^{-2\bar{\eta_1}x}}{1+e^{i(\theta_1+\bar{\theta}_1)}e^{4i(\eta_{1}^{2}-\bar{\eta_{1}}^{2})t}e^{-2(\eta_{1}+\bar{\eta_1})x}}\label{6a} ,\\
q^*(-x,t)=-\frac{2(\eta_{1}+\bar{\eta_1})e^{i\theta_1}e^{4i\eta_1^{2}t}e^{-2\eta_1x}}{1+e^{i(\theta_1+\bar{\theta}_1)}e^{4i(\eta_{1}^{2}-\bar{\eta_{1}}^{2})t}e^{-2(\eta_{1}+\bar{\eta_1})x}}\label{6b},
\end{eqnarray} \ees
from our one soliton solution (\ref{4}) for the following parametric choices, namely $k_{1}=i2\eta_1$, $\bar{k}_{1}=i2\bar{\eta}_1$, $\al_{1}=-2(\eta_1+\bar{\eta}_1)e^{i\bar{\theta}_{1}}$ and $\ba_{1}=-2(\eta_1+\bar{\eta}_1)e^{i\theta_{1}}$ (where $\eta_1$, $\bar{\eta}_1$, $\theta_1$ and $\bar{\theta}_1$, are all real). The above solution coincides with the one given in Ref. 3.

 In Ref. 1, the authors incorrectly claim that the more general soliton solutions obtained by us do not satisfy the $S$-symmetric equation (106) of their paper, which is same as Eq. (\ref{1}) given above. We point out here that our general soliton solution (\ref{4}) indeed satisfies the $S$-symmetric equation (106). We deduce the functions $q(x,t)$ and $q^*(-x,t)$ from (\ref{4}) for the non-singular soliton corresponding to Fig. 1 of our paper in Ref. 2, by fixing the parameters as $k_1=0.4+i$, $\bar{k}_1=-0.4+i$, $\al_1=1+i$, $\ba_1=1-i$, $\xi_1^{(0)}=\bar{\xi}_1^{(0)}=0$, that is 
\bea
q(x,t)=\frac{(1+i)e^{-(1+\frac{2i}{5})x+(\frac{4}{5}-\frac{21 i}{25})t}}{1+\frac{1}{2}e^{-2x+\frac{8}{5}t}},~
q^*(-x,t)=\frac{(1-i)e^{(-1+\frac{2i}{5})x+(\frac{4}{5}+\frac{21 i}{25})t}}{1+\frac{1}{2}e^{-2x+\frac{8}{5}t}}.
\label{5}
\eea
One can easily check that the above functions do satisfy the $S$-symmetric equation (106) given in Ref. 1 as well as each of Eqs. (\ref{1}) and (\ref{2}) of the present paper. This ensures that  the parameters chosen  by us in Ref. 2 for demonstrating the non-singualar one-soliton of reverse space NNLS Eq. (\ref{1}) is valid and correct one. 

 We also point out that G\"{u}rses and Pekcan in Ref. 1 wrongly calculated the functions $q(x,t)$ and $q^*(-x,t)$ from our general soliton solution (\ref{4}) for the parametric choice $k_1=0.4+i$, $\bar{k}_1=-0.4+i$, $\al_1=1+i$, $\ba_1=1-i$, $e^{\xi_1^{(0)}}=-1+i$ and $e^{\bar{\xi}_1^{(0)}}=1+i$ as 
\bea
q(x,t)=\frac{(2i)e^{-(1+\frac{2i}{5})x+(\frac{4}{5}-\frac{21 i}{25})t}}{1-e^{-2x+\frac{8}{5}t}},~~
q^*(-x,t)=\frac{(-2i)e^{(1-\frac{2i}{5})x+(\frac{4}{5}+\frac{21 i}{25})t}}{1-e^{2x+\frac{8}{5}t}}.\label{6}
\eea
However, the correct forms of $q(x,t)$ and $q^*(-x,t)$ deduced from (\ref{4}) for the above parametric choice are 
\bea
q(x,t)=\frac{(2i)e^{-(1+\frac{2i}{5})x+(\frac{4}{5}-\frac{21 i}{25})t}}{1-e^{-2x+\frac{8}{5}t}},~~
q^*(-x,t)=\frac{(2i)e^{(-1+\frac{2i}{5})x+(\frac{4}{5}+\frac{21 i}{25})t}}{1-e^{-2x+\frac{8}{5}t}}.\label{7}
\eea
 
It is evident that the wrong expressions given for the functions $q(x,t)$ and $q^*(-x,t)$ obviously do not satisfy the $S$-symmetric equation (106) given in Ref. 1, while our correct expressions (\ref{7}) given above do indeed satisfy it as well as Eqs. (1) and (2) of the present paper. These authors also claim that the parameters in our one and two general soliton solutions should obey the constraint equations (109) and (119) given in Ref. 1 which is obtained by imposing the relation (\ref{3}). As pointed out above   the $\cal{PT}$-symmetry broken one and two solutions of the reverse space NNLS Eq. (\ref{1})  need not obey the constraint equations (109) and (119) given in Ref. 1. 

 Finally, if one demands the condition (\ref{3}), for instance for the one soliton solution the parameters have to be constrained as $\al_1^*=\ba_1$, $k_1=\bar{k}_1^*$ and $\xi_1^{(0)}=\bar{\xi}_1^{(0)*}$ which corresponds to the $\cal{PT}$-symmetry unbroken case which  are satisfied by Eq. (106) of Ref. 2  or Eqs. (1) and (2) given above. For example, we deduce the functions $q(x,t)$ and $q^*(-x,t)$ for $k_1=0.4+i$, $\bar{k}_1=0.4-i$, $\al_1=1+i$, $\ba_1=1-i$, $\xi_1^{(0)}=0$ and $\bar{\xi}_1^{(0)}=0$ from (\ref{4}) in which the complex parameters obey the constraint Eq. (109) of Ref.  2 as
\bea
q(x,t)=\frac{(1+i)e^{(1+\frac{2i}{5})x+(\frac{4}{5}-\frac{21 i}{25})t}}{1-\frac{25}{8}e^{\frac{4i}{5}x+\frac{8}{5}t}},~
q^*(-x,t)=\frac{(1-i)e^{(-1+\frac{2i}{5})x+(\frac{4}{5}+\frac{21 i}{25})t}}{1-\frac{25}{8}e^{\frac{4i}{5}x+\frac{8}{5}t}}. \label{8}
\eea
The above functions also satisfy the $S$-symmetric NNLS equation (106) of Ref. 1 as well as Eqs. (1) and (2) given in the present paper. We also note that the above solution (\ref{8}) becomes singular at $x=\frac{5}{2}n\pi$ and $t=\frac{5}{8}\ln\frac{8}{25}$, $n$ is an integer, which is a generic property of the above type of reverse space NNLS equation, as pointed out by Ablowitz and Musslimani \cite{5}.

 Thus, to bring out both $\cal{PT}$-symmetry broken and unbroken soliton solutions of reverse space NNLS equation, one has to consider both Eqs. (\ref{1}) and (\ref{2}) simultaneously. The $\cal{PT}$-symmetry broken solution obtained by us need not satisfy the constraint equation (109) given in the recent paper of G\"{u}rses and Pekcan  \cite{2}. Consequently the parameters considered to demonstrate one- and two-soliton solutions in our paper Ref. 2 are valid ones and  they need not obey the constraint Eqs.(109) and (119) of Ref. 1 in general. 
\section*{Acknowledgements}
The work of MS forms part of a research project sponsored by DST-SERB, Government of India under the Grant No. EMR/2016/001818. The research work of ML is supported by a DST-SERB Distinguished Fellowship (ERB/F/6717/2017-18) and forms part of the DAE-NBHM research project (2/48 (5)/2015/NBHM (R.P.)/R\&D-II/14127).


\bibliography{mybibfile}

\end{document}